\documentclass
[preprint,prb,a4paper,showpacs,amsfonts,amssymb,final,unsortedaddress,fleqn,titlepage,12pt,floats,nobalancelastpage]{revtex4}%
\usepackage{amsfonts}
\usepackage{amsmath}
\usepackage{amssymb}
\usepackage{graphicx}%
\ifx\pdfoutput\relax\let\pdfoutput=\undefined\fi
\newcount\msipdfoutput
\ifx\pdfoutput\undefined\else
\ifcase\pdfoutput\else
\ifx\paperwidth\undefined\else
\ifdim\paperheight=0pt\relax\else\pdfpageheight\paperheight\fi
\ifdim\paperwidth=0pt\relax\else\pdfpagewidth\paperwidth\fi
\fi\fi\fi
\begin{document}
\author{M. ElMassalami and R. M. Saeed, C. M. Chaves}
\affiliation{Instituto de Fisica-UFRJ, CxP 68528, 21941-972, Rio de Janeiro,Brazil}
\author{H. Takeya,}
\affiliation{National Institute for Materials Science,1-2-1 Sengen, Tsukuba, Ibaraki,
305-0047, Japan}
\author{M. Doerr}
\affiliation{Institut f\"{u}r Festk\"{o}rperphysik, Technische Universit\"{a}t Dresden,
D-01062 Dresden, Germany}
\author{H. Michor}
\affiliation{Institut f\"{u}r Festk\"{o}rperphysik, Technische Universit\"{a}t Wien, A-1040
Wien, Austria}
\author{M. Rotter}
\affiliation{University of Oxford, Department of Physics Clarendon Laboratory, Parks Road
Oxford OX1 3PU}
\title{Specific heat in different magnetic phases of $R$Ni$_{2}$B$_{2}$C ($R$= Gd,
Ho, Er): theory and experiment }
\date{\today{}}

\begin{abstract}
The borocarbides $R$\textrm{Ni}$_{\mathrm{2}}$\textrm{B}$_{\mathrm{2}}%
$\textrm{C} ($R$=Gd, Ho, Er) exhibit a large variety of magnetic states and as
a consequence rich phase diagrams. We have analyzed the nature of these states
by specific heat investigations. The data were measured down to 0.5 K and up
to 80 kOe. The overall evolution of each $C_{mag}(T,H)$ curve is observed to
reflect faithfully the features of the corresponding $H-T$ phase diagram.
Within the lower ranges of temperature and fields, the calculations based on
linearized field-dependent spin-wave theory are found to
reproduce\ satisfactorily the measured $C_{mag}(T,H)$ curves: accordingly,
within these ranges, the thermodynamical properties of these compounds can be
rationalized in terms of only two parameters: the spin-wave energy gap and the
stiffness coefficient. For the intermediate fields ranges ($H_{1}<H<H_{sat}$)
wherein\ successive field-induced metamagnetic modes are stabilized, the
evolution of $C_{mag}(T,H)$ is discussed in terms of the Maxwell relation
$\left(  \partial C_{mag}/\partial H\right)  _{T}=T\left(  \partial
^{2}M/\partial T^{2}\right)  _{H}$. For the particular case of\ GdNi$_{2}%
$B$_{2}$C wherein the anisotropy is dictated by the classical dipole
interaction, $C_{mag}(T,H)$ across the whole ordered state is numerically
evaluated within the model of Jensen and Rotter [PRB \textbf{77 }(2008) 134408].

\end{abstract}
\maketitle

\section{Introduction}

The $H\mathrm{-}T$ magnetic phase diagrams of the heavy members of the
intermetallic $R$\textrm{Ni}$_{\text{2}}$\textrm{B}$_{\text{2}}$\textrm{C}
($R$= rare earth) magnets are characterized by a cascade of metamagnetic
transformations, higher Neel points $T_{N}$, and stronger saturation fields,
$H_{sat}$ (see Refs.1, 2 and references
therein\nocite{Muller01-interplay-review,Canfield-RNi2B2C-Hc2-review}). Both
$T_{N}$ and $H_{sat}$ are well scaled by de Gennes factors testifying to the
involvement of the indirect exchange coupling
mechanism.\cite{Cho96-RNi2B2C-deGeness} Moreover, except for Gd\textrm{Ni}%
$_{\text{2}}$\textrm{B}$_{\text{2}}$\textrm{C}, there are strong anisotropic
forces and the competition between these forces and the exchange interaction
is one of the main driving mechanism behind the characteristic features of
their magnetic phase
diagrams.\cite{Muller01-interplay-review,Canfield-RNi2B2C-Hc2-review} In
general, the role of these competing forces is analyzed in terms of a
Hamiltonian that consists of bilinear exchange, single-ion crystalline
electric field (CEF), and dipolar
interactions.\cite{Kalatsky98-clock-model,Amici-Thalmeier98-HoNi2B2C,Jensen02-mode-FM-ErNi2B2C}
Calculations based on such a Hamiltonian yield $H\mathrm{-}T$ phase diagrams
that are in reasonable agreement with experiments on, say, \textrm{HoNi}%
$_{\text{2}}$\textrm{B}$_{\text{2}}$\textrm{C}%
,\cite{Kalatsky98-clock-model,Amici-Thalmeier98-HoNi2B2C} \textrm{GdNi}%
$_{\text{2}}$\textrm{B}$_{\text{2}}$\textrm{C}%
,\cite{04-Rotter-Gd-based-MacPhase,Jensen08-GdNi2B2C-magnetoelstic-paradox},
and \textrm{ErNi}$_{\text{2}}$\textrm{B}$_{\text{2}}$\textrm{C.}%
\cite{Jensen02-mode-FM-ErNi2B2C}

The zero-field magnetic structures of the $R$\textrm{Ni}$_{\text{2}}%
$\textrm{B}$_{\text{2}}$\textrm{C} compounds can be divided into two broad
classes:\cite{Lynn97-RNi2B2C-ND-mag-crys-structure} (i) the
equal-amplitude\textit{,} collinear, and commensurate AFM structures ($R$=Pr,
Nd, Dy, Ho ): here the moments are coupled ferromagnetically within the layers
which are, in turn, antiferromagnetically stacked on each other along the
\textit{c} axis and (ii) the amplitude-modulated, incommensurate structures
($R$= Gd , Tb, Er, Tm); on lowering the temperature, these modulated
structures are gradually transformed into an equal-amplitude, squared-up
spin-density states.

On applying an external magnetic field along the easy-axis, both the
commensurate and incommensurate magnetic modes are observed to undergo a
cascade of magnetic phase transformations leading to a set of different
magnetic modes; each having a distinct propagation vector and a distinct total
magnetic moment. For the case of, say, \textrm{HoNi}$_{\text{2}}$%
\textrm{B}$_{\text{2}}$\textrm{C} and \textrm{ErNi}$_{\text{2}}$%
\textrm{B}$_{\text{2}}$\textrm{C,} there emerges a very rich and interesting
$H$-$T$ magnetic phase
diagram\cite{Campbell00-Er-HT-diagram,Campbell-Ho-HT-Diagram,Detlefs00-Ho-Field-neutronDiff}
with characteristic features that are reflected in all measured
thermodynamical properties. As an example, the low-temperature magnetization
isotherms show a cascade of step-like metamagnetic phase transformations with
constant plateau between the consecutive
transformations.\cite{Canfield-HoNi2B2C-M-Cp,Canfield96-Er-WF-HT-diagram,Canfield97-Ho-angularMetamagnet,Budko00-ErNi2B2C-angularHTdiagram,02-ErNi2B2C-magnetostriction}
As the low-temperature magnetic properties are governed by the excitation
spectra of the magnon quasiparticles then, according to the spin-wave theory,
the surge of such a stair-like features is a manifestation of related
field-dependent distinct features in the dispersion relations or density of
states of the magnon quasiparticles. Then a detailed spin-wave analysis of the
$H$- and $T$-dependence of the thermodynamical properties would contribute
positively to the understanding of the magnetic phase diagrams and the related
thermodynamical properties of these borocarbide magnets. Using field-dependent
specific heat measurements on single-crystals of $R$\textrm{Ni}$_{\text{2}}%
$\textrm{B}$_{\text{2}}$\textrm{C} ($R$ = Gd, Ho, Er), this work investigated
the magnetic contribution to the specific heat when $T$ and $H$ are varied
across these $H$-$T$ phase diagrams.

The magnetic structures of the compounds under study are well established for
the high- and low-field part of the phase diagram: within the
intermediate-field range, the structures are either unknown or too complex.
Furthermore, it happened that the highest magnetic field available for this
study is lower than the one needed to reach the high-field saturated range.
Accordingly, the comparison of the measured and calculated magnetic specific
heats\ (the latter is based on spin-wave theory) was carried out only for the
lower field limit (\S \ II.A and IV.C). For the intermediate-field\ range (as
well as across the available $H-T$ plane of the phase diagram), basic
thermodynamical relations were employed to relate the field evolution of the
measured magnetic specific heat to the thermal evolution of the isofield
magnetization curves (\S \ II.C and \S \ IV.C). Finally, since the linearized
magnon theory is applicable only for the lower temperature range, mean-field
model calculations were carried out to evaluate the magnetic specific heat
within the higher range of the ordered state; this part of the study was
restricted to only \textrm{GdNi}$_{\text{2}}$\textrm{B}$_{\text{2}}$\textrm{C}
(\S \ II.B and \S \ IV.C.1.b): in contrast to $R$ = Ho and Er,\ \textrm{GdNi}%
$_{\text{2}}$\textrm{B}$_{\text{2}}$\textrm{C} has negligible crystalline
field anisotropy and as such the classical dipole interactions play a crucial
role in shaping its magnetic structure.

\section{Theoretical Background}

\subsection{Field-dependent magnon specific heat of $R$\textrm{Ni}$_{\text{2}%
}$\textrm{B}$_{\text{2}}$\textrm{C}.}

The magnetic properties of $R$\textrm{Ni}$_{\text{2}}$\textrm{B}$_{\text{2}}%
$\textrm{C} are usually described in terms of the Hamiltonian:%
\begin{equation}
\mathcal{H}=\mathcal{H}_{Z}+\mathcal{H}_{ex}+\mathcal{H}_{CEF}\text{,}%
\label{Eq-Htot}%
\end{equation}
where $\mathcal{H}_{Z}$ is the Zeeman term with the field $H$ applied at angle
$\theta$ away from the easy axis:%
\begin{equation}
\mathcal{H}_{Z}=-g\mu_{B}\vec{H}.\sum_{j}\overrightarrow{S_{j}},\label{Eq-Hz}%
\end{equation}
where $\overrightarrow{S_{j}}$ is the total angular momentum operator of the
$j^{\mathrm{th}}$ $R^{3+}$-ion, $g_{J}$ is the Land\'{e} factor, and $\mu_{B}$
is the Bohr magneton. The bilinear isotropic exchange interaction is:%

\begin{equation}
\mathcal{H}_{ex}=-\frac{1}{2}\sum_{ij}\mathcal{J}(ij)\overrightarrow{S_{i}%
}.\overrightarrow{S_{j}}, \label{Eq-Hex}%
\end{equation}
where $\mathcal{J}(ij)$ is the isotropic coupling between $\overrightarrow
{S}_{i}$ and $\overrightarrow{S_{j}}$. Considering the low-temperature
commensurate AFM\ structures of $R$\textrm{Ni}$_{\text{2}}$\textrm{B}%
$_{\text{2}}$\textrm{C} then $\mathcal{H}_{ex}$ (Eq. \ref{Eq-Hex}) can be
split into two terms:
\begin{equation}
\mathcal{H}_{ex}=+\sum_{<ij>,L\in A,B}\mathcal{J}_{ij}^{L}\overrightarrow
{S_{i}^{L}}.\overrightarrow{S_{j}^{L}}+\sum_{<ij>A,B}\mathcal{J}_{ij}%
^{AB}\overrightarrow{S_{i}^{A}}.\overrightarrow{S_{j}^{B}}
\label{Eq-Hexlayered}%
\end{equation}
$\mathcal{J}_{ij}^{A}$ couples moments $\overrightarrow{S_{i}^{L}}$ and
$\overrightarrow{S_{j}^{L}}$ within the same layer ($L=$ $A$ or $B$) and
$\mathcal{J}_{ij}^{AB}$ couples moments from different layers. At lower
temperatures, we assume that the net effect of $\mathcal{H}_{CEF}$ on the
magnetic moment can be approximated by an easy-axis anisotropic field
(represented by $\vec{H}_{a}$),\cite{Kittel-SpinWave,Joenk62-Cm-field}
consequently:%
\begin{equation}
\mathcal{H}_{CEF}+\mathcal{H}_{Z}=-g\mu_{B}(\vec{H}_{a}+\vec{H}).\sum_{i\in
A}\overrightarrow{S_{i}^{A}}-g\mu_{B}(\vec{H}_{a}-\vec{H}).\sum_{j\in
B}\overrightarrow{S_{j}^{B}}. \label{Eq-Hcef}%
\end{equation}

Let us consider the following useful simplification: as far as the magnon
propagation is concerned, the low-temperature squared-up incommensurate
magnetic structures is assumed to behave as if it is a collinear AFM structure
(for, say, \textrm{ErNi}$_{\text{2}}$\textrm{B}$_{\text{2}}$\textrm{C}, this
ignores the presence of the kinks along the $a$ axis):\ such a simplification
is motivated by the remarkable feature that though $R$\textrm{Ni}$_{\text{2}}%
$\textrm{B}$_{\text{2}}$\textrm{C }( $R$\textrm{=Ho, Er}) have different
magnetic structures,\cite{Lynn97-RNi2B2C-ND-mag-crys-structure} however they
manifest similar stair-like features in their $M(T,H)$
isotherms\cite{Canfield-HoNi2B2C-M-Cp,Canfield96-Er-WF-HT-diagram,Canfield97-Ho-angularMetamagnet,Budko00-ErNi2B2C-angularHTdiagram,02-ErNi2B2C-magnetostriction}
as well as similar expression for their zero-field $C_{mag}(T,H)$
curves.\cite{03-Magnon-RNi2B2C}

Using standard linearized spin-wave procedures,\cite{Kittel-SpinWave} the
total Hamiltonian (Eq. \ref{Eq-Htot}) is diagonalized and consequently the
magnetic specific heat is calculated. Based on the strength of the applied
field, three different situations can be distinguished: (i) the low-field
limit ($H<H_{\text{sf}}$: the spin-flop field) wherein the spins are basically
aligned along the preferred orientation, (ii) the intermediate field regime
wherein a series of field-induced metamagnetic states are stabilized. and
(iii)the high field limit ($H\geqslant H_{sat}$: saturated field) wherein
spins are driven towards paramagnetic saturation.

\subsubsection{Low-field approximation $(H<H_{\text{sf}})$}

The dispersion relation of the two modes (+ and -) are:
\begin{align}
\hbar\omega_{k}^{\pm}  &  =+\sqrt{A^{2}-B^{2}}\pm g\mu_{B}Hcos(\theta
),\nonumber\\
A  &  =4S\mathcal{J}_{o}\left\{  1-0.5\left[  \cos(ak_{x})+\cos(ak_{y}%
)\right]  \right\}  +8S\mathcal{J}_{1}+g\mu_{B}H_{a},\nonumber\\
B  &  =2S\mathcal{J}_{1}\left\{  \cos(\frac{a}{2}k_{x}+\frac{a}{2}k_{y}%
+\frac{c}{2}k_{z})+\cos(\frac{a}{2}k_{x}+\frac{a}{2}k_{y}-\frac{c}{2}%
k_{z})\right. \nonumber\\
&  \left.  +\cos(\frac{a}{2}k_{x}-\frac{a}{2}k_{y}+\frac{c}{2}k_{z}%
)+\cos(-\frac{a}{2}k_{x}+\frac{a}{2}k_{y}+\frac{c}{2}k_{z})\right\}
\label{Eq-disp}%
\end{align}
where $\mathcal{J}_{0}$ and $\mathcal{J}_{1}$ are effective exchange couplings
among the nearest neighboring within, respectively, the same plane and within
neighboring planes. Accordingly, there are two distinct field-dependent energy
gaps:
\begin{equation}
\Delta_{a}^{\pm}(H)=+\sqrt{\left(  g\mu_{B}H_{a}\right)  ^{2}+16\mathcal{J}%
_{1}g\mu_{B}SH_{a}}\pm g\mu_{B}Hcos(\theta)=\Delta_{a}\pm g\mu_{B}%
Hcos(\theta); \label{Eq-gap-AFM}%
\end{equation}
evidently $\Delta^{\pm}(H)$ do not depend on $\mathcal{J}_{o}$ and their
separation is linearly related to $Hcos(\theta)$.

At lower temperature where long-wave limit is valid, the integration over the
allowed $k$-space yields the magnon specific heat:%
\begin{align}
C_{mag}(T,H)  &  =R\Delta_{a}^{4}/(4\pi^{2}D_{a}^{3}T^{2})\sum_{m=1}^{\infty
}\left\{  T\cosh(m\xi/T)\left[  K_{4}(m\Delta_{a}/T)+\left\langle 1+2\xi
^{2}/\Delta_{a}^{2}\right\rangle K_{2}(m\Delta_{a}/T)\right]  \right.
\nonumber\\
&  \left.  -0.5m\xi\sinh(m\xi/T)\left[  K_{4}(m\Delta_{a}/T)-K_{0}(m\Delta
_{a}/T)\right]  \right\}  \label{Eq-Cm-AFM-lowH}%
\end{align}
where $K_{n}(m\Delta_{a}/T)$ represents the modified Bessel function of the
second kind, $\xi=g\mu_{B}Hcos(\theta)/k_{B}$ and $D_{a}$ is a measure of the
stiffness and is function of the exchange couplings and magnetic anisotropy:%
\begin{equation}
D_{a}=\left(  16(\mathcal{J}_{0}+\mathcal{J}_{1})\mathcal{J}_{0}%
S^{2}+2\mathcal{J}_{0}S(g\mu_{B}H_{a})\right)  ^{\frac{1}{3}}.\left(
4\ \mathcal{J}_{1}S\right)  ^{\frac{1}{3}} \label{Eq-Stiff-AFM}%
\end{equation}

Due to the type of the undertaken approximations, Eq. \ref{Eq-Cm-AFM-lowH}
does not hold for $H\cos(\theta)$ equal or higher than the spin-flop field
$H_{sf}$ which is the value at which the lower branch goes to zero. The field
influence on Eq. \ref{Eq-Cm-AFM-lowH} enters only through\ $Hcos(\theta)$.
Furthermore, this equation reduces to Eq. 25 of Joenk\cite{Joenk62-Cm-field}
when $\mathcal{J}_{0}=\mathcal{J}_{1}$ and it gives the well-known $T^{3}$
relation when $k_{B}T$ $>\Delta_{a}$.\cite{Kittel-SpinWave}

For $T\rightarrow0$, $\Delta_{a}/T\rightarrow\infty,$ and $\xi<\Delta_{a}$,
Eq. \ref{Eq-Cm-AFM-lowH} reduces to:%
\begin{equation}
C_{mag}(T,H)\simeq\frac{R\Delta_{a}^{7/2}}{2^{1/2}\pi^{3/2}D_{a}^{3}T^{1/2}%
}\exp(\frac{\xi-\Delta_{a}}{T})\left[  1-2\frac{\xi}{\Delta_{a}}+\left(
\frac{\xi}{\Delta_{a}}\right)  ^{2}..\right]  , \label{Eq-low-T-limit}%
\end{equation}
indicating a dominant exponential character within this $H-T$ region.

\subsubsection{Intermediate fields range ($H_{sf}<H<H_{sat})$}

As mentioned in \S \ I, the absence of a detailed description of the involved
magnetic structures together with the absence of analytical expression for the
dispersion relations hinder any direct evaluation of the magnon specific heat
contribution within this range.

\subsubsection{The high-field limit ($H>H_{sat}$)}

Due to the limitation of our experimental conditions, we were not able to
probe the magnon contribution within the region of the\ $H-T$ phase diagram
wherein the induced FM\ state is established. Nevertheless for completeness
sake we derive the magnetic contribution for this state in appendix A below.

\subsection{Mean-field Model calculation of $C_{mag}(T,H)$ of
GdNi$_{\mathrm{2}}$B$_{\mathrm{2}}$C}

The above calculations of the magnon $C_{mag}(T,H)$ of $R$\textrm{Ni}%
$_{\mathrm{2}}$\textrm{B}$_{\mathrm{2}}$\textrm{C} compounds are valid only
within the lower temperature range, mostly below liquid helium temperatures.
To investigate the magnetic contribution within the whole magnetically ordered
range, we resort to mean-field model calculations applied on the simpler case
of the ordered $^{7/2}S$ -moments of \textrm{GdNi}$_{\mathrm{2}}$%
\textrm{B}$_{\mathrm{2}}$\textrm{C}. Jensen and
Rotter\cite{Jensen08-GdNi2B2C-magnetoelstic-paradox} showed that a model
consisting of\ a sum of the bilinear Heisenberg exchange term and the
classical dipole interaction is able to explain the features of the whole
magnetic phase diagram as well as the so-called magnetoelastic
paradox.\cite{06-Gd-paradox} and the zero-field $C_{mag}(T)$ of \textrm{GdNi}%
$_{\mathrm{2}}$\textrm{B}$_{\mathrm{2}}$\textrm{C}. here, in this work, their
calculations of  $C_{mag}(T,H)$ are extended to fields up to $H\leq$ 80 kOe
using the McPhase program package
(www.mcphase.de).\cite{rotter04-Mcphase-program}

\subsection{Application of the Maxwell relation to correlate $M(T,H)$\ and
$C_{mag}(T,H)$}

An applied field on an AFM\ mode tends to remove the degeneracy appearing in
Eqs. \ref{Eq-disp} and \ref{Eq-gap-AFM}. As a consequence, the fractional
contribution of the lower mode to $C_{mag}(T,H)$ increases leading to $\left(
\partial C_{mag}/\partial H\right)  _{T}>0$. On the other hand, for
$H\rightarrow$ $H_{sat}$, an increase in $H$ would induce a gradual decrease
in $C_{mag}(T,H)$ resulting in $\left(  \partial C_{mag}/\partial H\right)
_{T}<0.$ Similar arguments hold for $M(T,H)$. In fact the following Maxwell
equations relate the evolution of $M(T,H)$\ to that of $C_{mag}(T,H)$:%
\begin{subequations}
\begin{align}
\left(  \partial S_{mag}/\partial H\right)  _{T} &  =\left(  \partial
M/\partial T\right)  _{H}\label{Eq-C-M-relations(a)}\\
\left(  \partial C_{mag}/\partial H\right)  _{T} &  =T\left(  \partial
^{2}M/\partial T^{2}\right)  _{H}\text{.}\label{Eq-C-M-relations(b)}%
\end{align}

These general relations are very helpful in the analysis of $S_{mag}(T,H)$ and
$C_{mag}(T,H)$ since it is much easier to measure the thermal evolution of
$M(T,H)$ than the field dependence of $C_{mag}(T,H)$. Furthermore, they hold
across the whole $H-T$ phase diagram and are independent of field orientation
(whether $H\Vert a$ or $H\Vert c$) or the type of the involved magnetic
structure (whether commensurate or incommensurate). This utility is most
welcomed when investigating those regions of the $H-T$ phase diagrams wherein
the spin-wave analysis of \S \ II.A is not applicable.

Evidently each of $C_{mag}(T<T_{N},H)$ and $M(T<T_{N},H)$ would mirror the
exotic features of the $H-T$ phase diagrams of $R$\textrm{Ni}$_{\text{2}}%
$\textrm{B}$_{\text{2}}$\textrm{C}%
;\cite{Canfield-HoNi2B2C-M-Cp,Canfield96-Er-WF-HT-diagram,Canfield97-Ho-angularMetamagnet,Budko00-ErNi2B2C-angularHTdiagram,Budko00-ErNi2B2C-angularHTdiagram}
in fact, these diagrams have been constructed from the magnetic anomalies
occurring in $C_{mag}(T<T_{N},H)$ and $M(T<T_{N},H)$ as well as many other
magnetic properties such as the integrated intensities of the magnetic neutron
diffractograms;\cite{Campbell00-Er-HT-diagram,Campbell-Ho-HT-Diagram} the
magnetostriction,\cite{02-ErNi2B2C-magnetostriction,03-GdNi2B2C-singlecrystal-1}
and the
magnetoresistivity.\cite{96-HoNi2B2C-Mag-resis,Fisher97-RNi2B2C-magnetoresistivity}
Here in this work, we use Eqs. \ref{Eq-C-M-relations(a)} and
\ref{Eq-C-M-relations(b)} to discuss the general trend of $C_{mag}(T,H)$ in
terms of the trend of the extensively reported $M(T,H)$%
\ curves.\cite{Muller01-interplay-review}

\section{Experimental}

The $R$\textrm{Ni}$_{\text{2}}$\textrm{B}$_{\text{2}}$\textrm{C} ($R$= Gd, Ho,
Er) single-crystals were selected for this study because they offer a good
representation of the magnetic properties of the whole$\ R$\textrm{Ni}%
$_{\text{2}}$\textrm{B}$_{\text{2}}$\textrm{C} series: \textrm{HoNi}%
$_{\text{2}}$\textrm{B}$_{\text{2}}$\textrm{C} is a typical representative of
the collinear, commensurate, AFM\ structures while \textrm{ErNi}$_{\text{2}}%
$\textrm{B}$_{\text{2}}$\textrm{C} and \textrm{GdNi}$_{\text{2}}$%
\textrm{B}$_{\text{2}}$\textrm{C} are good representatives of the modulated,
incommensurate magnetic structures; the anisotropic forces are stronger in the
former\cite{Cho95-ErNi2B2C-ansitropy} while extremely weaker in the
latter.\cite{Canfield-GdNi2B2C,03-GdNi2B2C-singlecrystal-1} \textrm{GdNi}%
$_{\text{2}}$\textrm{B}$_{\text{2}}$\textrm{C} has negligible CEF forces;
nonetheless, it is shown that the dipolar forces are essential for directing
the moments transversal to the propagation vector
(0.55,0,0).\cite{04-Rotter-Gd-based-MacPhase}

The single crystals of these three representatives, together with that of the
reference \textrm{YNi}$_{\text{2}}$\textrm{B}$_{\text{2}}$\textrm{C}, were
grown by floating zone method.\cite{Takeya96a-FZ-method} Results from
extensive structural and physical characterizations are in good agreement with
the reported data, confirming the good quality of our crystals. The
temperature-dependent specific heat at fixed fields was measured on two
different setups. One is pulse-type adiabatic calorimeter [500 mK
$<$%
$T$%
$<$%
25 K, 120 kOe] and the second is a quasi-adiabatic setup with a temperature
range covering 1.5-100K and a field up to 80 kOe. For the sake of
completeness, we quote some of our results (in particular the zero-field
magnetic specific heat) that had been reported in Refs.
\cite{03-AFSup-RNi2B2C-Cm,03-Magnon-RNi2B2C,04-Magnetocaloric}.

\section{\textbf{Results and Discussion}}

\subsection{Electronic and lattice contribution}

For all compounds, the total specific heat $C_{tot}$ was analyzed as a sum of
an electronic $C_{e}$, a phonon $C_{ph}$, a nuclear $C_{N}$, and a magnetic
contribution $C_{mag}$:\
\end{subequations}
\begin{equation}
C_{tot}=C_{e}+C_{ph}+C_{n}+C_{mag}\label{Eq-Ctot}%
\end{equation}
For consistency reasons, the lattice contribution was estimated from the
specific heat of the isomorphous single-crystal of \textrm{YNi}$_{\text{2}}%
$\textrm{B}$_{\text{2}}$\textrm{C}.\cite{03-AFSup-RNi2B2C-Cm} We tried other
means of estimating the phonon contribution (such as mass
normalization\cite{Bouvier-Gd-amplitude-modulated} or using other nonmagnetic
isomorphs\cite{Michor95-C-RNi2B2C,03-AFSup-RNi2B2C-Cm}). It is found out that
the spin-wave fit-parameters based on different methods are differing by only
a few percent; other than this variation, the use of different estimation
process does not influence the conclusions drawn from this work.

The low-temperature, normal-state electronic contribution of these compounds
was evaluated from $C_{tot}(T,H)$ of \textrm{YNi}$_{\text{2}}$\textrm{B}%
$_{\text{2}}$\textrm{C} single-crystal (the same as that used in Ref
.\cite{03-AFSup-RNi2B2C-Cm}). Within the normal state (or $H>H_{c2}$ if
applicable), the electronic, phonic, or nuclear contributions are taken to be
field independent (see below). An illustration of the various contributions of
$R$\textrm{Ni}$_{\text{2}}$\textrm{B}$_{\text{2}}$\textrm{C} ($R$\ =Ho, Er)
are shown in Fig. \ref{Fig-Ho-Er-Cn}; evidently for all$\ R$= Gd, Ho, Er
compounds, the diamagnetic contribution is small in comparison with the
magnetic or nuclear term.%
\begin{figure}
[th]
\begin{center}
\includegraphics[
natheight=4.687300in,
natwidth=3.207600in,
height=4.6873in,
width=3.2076in
]%
{D:/Spin-Wave/graphics/RNi2B2C-Field-Spheat-Nuc-Ho-Er__1.pdf}%
\caption{(Color online) Log-log plot of the zero-field $C_{tot}(T)$ versus $T$
showing the various individual contributions that are contained in Eq.
\ref{Eq-Ctot}: (a) HoNi$_{2}$B$_{2}$C and (b) ErNi$_{2}$B$_{2}$C. The
triangles (circles) denote the experimental total (magnetic) specific heats,
while the lines represent the various calculated contributions: dotted,
dashed, solid-thin (red) and solid-thick (black) lines represent,
respectively, the electron+phonon, nuclear, magnetic, and total contribution.
Both nuclear and magnetic contributions are obtained from the least-square
fits (see text). }%
\label{Fig-Ho-Er-Cn}%
\end{center}
\end{figure}

\subsection{Nuclear Contribution}

The magnetic contribution was obtained\ as follows: after subtracting the
electronic and phonon by the process explained in \S IV.A, the resultant
($C_{tot}-C_{e}-C_{ph}$) is confronted with the sum of the nuclear (Eq.
\ref{Eq-Cn}) and magnetic (Eq. \ref{Eq-Cm-AFM-lowH}) terms. Finally, the
magnetic contribution is obtained after subtracting out the nuclear specific
heat term which\ is usually given
as:\cite{Kruis-Nclr-Ho-Tb-Schottky,Kruis-Nclr-Eu-Er-Schottky}%

\begin{align}
C_{n}(T) &  =\Lambda_{iso}\left(
{\displaystyle\sum\limits_{i=-I}^{+I}}
{\displaystyle\sum\limits_{j=-I}^{+I}}
\left(  \omega_{j}^{2}-\omega_{i}\omega_{j}\right)  \exp(\frac{-\omega
_{j}-\omega_{i}}{k_{B}T})\right)  /\left[
{\displaystyle\sum\limits_{i=-I}^{+I}}
{\displaystyle\sum\limits_{j=-I}^{+I}}
\exp(\frac{-\omega_{j}-\omega_{i}}{k_{B}T})\right]  ,\nonumber\\
\omega_{i} &  =(\alpha_{int}+\alpha_{ext}).i-P.(i^{2}-\frac{I(I+1)}%
{3})\label{Eq-Cn}%
\end{align}
where $\Lambda_{iso}$ is the isotope abundance, $\alpha_{int}$ and $P$ are the
magnetic dipole and electric quadrupole interaction parameters of the nuclear
spins, respectively; $I$ is the total nuclear spin while $i$ is its component
along the quantization axis. $\alpha_{ext}$ is associated with the externally
applied magnetic field which is extremely small if compared to the internal
field: for, say, \textrm{ErNi}$_{\text{2}}$\textrm{B}$_{\text{2}}$%
\textrm{C},\cite{Kruis-Nclr-Eu-Er-Schottky} $H_{int}$ $\approx$7x10$^{6}$ Oe
and thus the highest applied field is only $\sim$1\% of $H_{int}%
$.\begin{table}[th]
\caption{The nuclear hyperfine parameters ($\Lambda_{iso}$, $\alpha_{int}$,
and $P$) and the spin-wave parameters ($\Delta_{a}$ and $D_{a}$ ) of
HoNi$_{2}$B$_{2}$C and ErNi$_{2}$B$_{2}$C which are obtained after fitting the
experimental ($C_{tot}-C_{e}-C_{ph}$) curves to the sum of the nuclear (Eq.
\ref{Eq-Cn}) and magnetic (Eq. \ref{Eq-Cm-AFM-lowH}) terms. The specified
temperature range indicates the region wherein the nuclear contribution is
dominant (for the temperature range of the fit see \S \ IV.C.2 and
\S \ IVC.3). The obtained nuclear parameters are compared with the
corresponding parameters of the $R$-metal (see text).}%
\begin{tabular}
[c]{cccccccc}\hline\hline
$R$Ni$_{2}$B$_{2}$C & Temp. range & $\Lambda_{iso}$ & I & $\alpha_{int}$ & $P$
& $\Delta$ & $D$\\\hline
& (K) & \% &  & K & mK & K & K\\\hline
HoNi$_{2}$B$_{2}$C & [0.5,1.5] & 100 & 7/2 & 0.362 & -9.6 & $7.7\pm0.3$ &
4.6$\pm0.2$\\
Ho metal\cite{Kruis-Nclr-Ho-Tb-Schottky} & [0.03,0.5] & 100 & 7/2 & 0.32 &
7.0 &  & \\
ErNi$_{2}$B$_{2}$C & [0.1,0.5] & 90(5) & 7/2 & 0.054 & 2.9 & $7.0\pm0.1$ &
3.0$\pm0.1$\\
Er metal\cite{Kruis-Nclr-Eu-Er-Schottky} & [0.03,0.8] & 100 & 7/2 & 0.042 &
-2.7 &  & \\\hline\hline
\end{tabular}
\label{Tab-Nuclear}%
\end{table}

The least square fit involves the simultaneous search for the best values of
the five parameters, namely the nuclear parameters $\Lambda_{iso}$,
$\alpha_{int}$, and $P$ as well as the spin-wave parameters $\Delta_{a}$ and
$D_{a}$ (see below). After the substitution of the obtained nuclear parameters
(Table \ref{Tab-Nuclear}), the expression of $C_{n}(T)$ (see Eq. \ref{Eq-Cn})
does reproduce satisfactorily the measured nuclear specific heat of
\textrm{ErNi}$_{\text{2}}$\textrm{B}$_{\text{2}}$\textrm{C} and \textrm{HoNi}%
$_{\text{2}}$\textrm{B}$_{\text{2}}$\textrm{C} (\textrm{GdNi}$_{\text{2}}%
$\textrm{B}$_{\text{2}}$\textrm{C} has no nuclear contribution): the overall
fits are shown in Fig. \ref{Fig-Ho-Er-Cn}. It is evident that $\alpha_{int}$
of the studied $R^{3+}$ compounds are close to the values reported for the
corresponding rare-earth
metals\cite{Kruis-Nclr-Ho-Tb-Schottky,Kruis-Nclr-Eu-Er-Schottky} indicating
that the hyperfine field is determined mainly by the internal electronic
configuration of the $R^{3+}$ ion. On the other hand, the $P$ parameters are
extremely small; this is not surprising since\ the point group of the sites at
which the $R^{3+}$ nucleus resides is $D_{4h}$ in borocarbides and $D_{6h}$ in
the elemental rare earth.

\subsection{Magnetic contribution}

Based on the general features of the $C_{mag}(T,H)$ curves (shown in Figs.
\ref{Fig-Gd-Ctot}, \ref{Fig-Ho-Ctot}, and \ref{Fig-Er-Ctot}), one
distinguishes four temperature regions: (i) a paramagnetic region,
$T>T_{N}(H)$, wherein $C_{mag}(T,H)$ is due to change in the population of the
crystal field levels, (ii) a critical region, $T\approx T_{N}(H)$, wherein
$C_{mag}(T,H)$ is related to critical phenomena, (iii) an intermediate region,
$T_{X}<T<T_{N}$ ($T_{X}=$ $T_{R}$ for \textrm{GdNi}$_{\text{2}}$%
\textrm{B}$_{\text{2}}$\textrm{C} or $T_{WFM}$ for \textrm{ErNi}$_{\text{2}}%
$\textrm{B}$_{\text{2}}$\textrm{C}) which encompasses the sine modulated
states. Within this region, the spin-wave analysis of \S \ II.A.1-2 is not
applicable. Finally (iv) the low--temperature region which should be
restricted if one intends to analyze $C_{mag}(T,H_{\Vert\theta})$\ in terms of
the linearized magnon theory: below we restrict this range to $T<$ $\Delta$
and $H<H_{\text{sf}}$ (none of the compounds under study presents the
$H>H_{sat}$ case). Within this region, the collinear AFM/squared-up states are
established and the measured $C_{mag}(T,H)$ is to be confronted with Eq.
\ref{Eq-Cm-AFM-lowH}. It is worth mentioning that for extremely
low-temperatures ($T<<\Delta$) , the exponential decaying character of
$C_{mag}(T,H)$ (see Eq. \ref{Eq-low-T-limit}) and the relatively large
contribution of the nuclear Schottky contribution limit the usefulness of the
lower temperature range for the magnon analysis.

\subsubsection{\textrm{GdNi}$_{\text{2}}$\textrm{B}$_{\text{2}}$\textrm{C}}

The zero-field magnetic structure of \textrm{GdNi}$_{\text{2}}$\textrm{B}%
$_{\text{2}}$\textrm{C} (\textit{T}$_{N}=19.5$ K) is an incommensurate
sine-modulated structure\ (moments along $b$ axis and $\overrightarrow{q}%
=$0.551$a^{\ast}$%
).\cite{Canfield-GdNi2B2C,Detlefs96-GdNi2B2C-XRES,Tomala98-GdNi2B2C-MES} At
$T_{R}\approx$13.5 K, a moment reorientation sets-in leading to an additional
modulated mode transversely polarized and having a small amplitude along the
$c$-axis.\cite{Detlefs96-GdNi2B2C-XRES,Tomala98-GdNi2B2C-MES} Two $H-T\ $phase
diagrams were reported:\cite{03-GdNi2B2C-singlecrystal-1} one for $H\parallel
a$ and another for $H\parallel c$. The phase diagram for $H\mathrm{\parallel
}a$ shows three field-induced magnetic phase transitions [see inset of
Fig.\ref{Fig-Gd-Ctot} (a)]: (i) the saturation boundary with $H_{sat}$; (ii)
the reorientation boundary $H_{R}^{\mathrm{\parallel}a}$; and finally (iii)
the domain-wall boundary $H_{D}^{\mathrm{\parallel}a}$. In contrast, the phase
diagram for $H\mathrm{\parallel}c$ shows only two transitions [see inset of
Fig.\ref{Fig-Gd-Ctot} (b)]: $H_{sat}^{\mathrm{\parallel}c}$ and $H_{R}%
^{\mathrm{\parallel}c}$; in comparison, the thermal evolution of the the
former (latter) is similar to (different from) that of $H_{sat}%
^{\mathrm{\parallel}a}$\ ($H_{R}^{\mathrm{\parallel}a}$). None of the two
phase diagrams shows those characteristic $H$-induced cascade of metamagnetic
phase transitions which are common in, say, the case of $R$ = Er, Ho; this is
attributed to the absence of strong anisotropic features.

As mentioned above, the experimental magnetic specific heat of \textrm{GdNi}%
$_{\text{2}}$\textrm{B}$_{\text{2}}$\textrm{C} would be confronted with two
model calculations: (i) the magnon calculation which is valid for sub-helium
temperature (\S \ II.A) and the model calculations of Jensen and
Rotter\cite{Jensen08-GdNi2B2C-magnetoelstic-paradox} which are extended from
the helium-temperature range up to $T_{N}$ (\S \ II.B).
\begin{figure}
[th]
\begin{center}
\includegraphics[
natheight=4.370800in,
natwidth=3.262900in,
height=4.3708in,
width=3.2629in
]%
{D:/Spin-Wave/graphics/RNi2B2C-Field-Spheat-Gd-tot__2.pdf}%
\caption{(Color online) The isofield $C_{mag}$($T,H$) of \textrm{GdNi}$_{2}%
$\textrm{B}$_{2}$\textrm{C} at various magnetic fields applied along (a)
$H\mathrm{\parallel}a$ and (b) $H\mathrm{\parallel}c$. The inset at the
bottom-left (top-left) illustrates the $H-T$ phase diagram for
$H\mathrm{\parallel}a$ ($H\mathrm{\parallel}c$);
\cite{03-GdNi2B2C-singlecrystal-1} there, the horizontal dotted lines
represent the fields that were applied during these measurements. Evidently,
both $H_{R}^{\mathrm{\parallel}c}$ and $H_{R}^{\mathrm{\parallel}c}$ anomalies
in the $C_{mag}$($T,H$) curve do reproduce the boundaries of the phase
diagrams, in particular the reentrant feature appearing for the
$H\mathrm{\parallel}a$ case (see text).}%
\label{Fig-Gd-Ctot}%
\end{center}
\end{figure}

\paragraph{Magnon contribution to $C_{mag}(T,H)$}

Figure \ref{Fig-Gd-Ctot} shows that the most prominent features of
$C_{mag}(T,H)$ of \textrm{GdNi}$_{\text{2}}$\textrm{B}$_{\text{2}}$\textrm{C}
are (i) the characteristic and distinct evolution of the $H_{sat}(T)$ and
$H_{R}(T)$ curves and that (ii) within a certain region of $T$ and $H$,
$C_{mag}(T,H)$ appears to be hardly influence by $H.$ The cause of this
apparent collapse of the $C_{mag}$($T,H$) curves becomes clear if we compare
these curves with the predictions of Eq. \ref{Eq-Cm-AFM-lowH}: Fig.
\ref{Fig-Gd-Cfit} shows a fit of the measured $C_{mag}$($T,H=0$) curve to Eq.
\ref{Eq-Cm-AFM-lowH} and the obtained fit parameters are $\Delta=2.9\pm0.1$ K
and $D=$5.6$\pm0.1$ K which are close but, due to the difference in the
temperature range, are better than those reported in Ref.
\cite{03-Magnon-RNi2B2C}. As both $\Delta_{a}$ and $D_{a}$ are field
independent, then the insertion of these values into Eq. \ref{Eq-Cm-AFM-lowH}%
\ leads to the calculated $C_{mag}$($T,H$) curves (with no adjustable
parameters) for all fields up to $H<k_{B}\Delta_{a}/(g\mu_{B})\sim20$ kOe
which is the field above or equal to which the magnon calculations based on
Eq. \ref{Eq-Cm-AFM-lowH} are not valid. Evidently the calculated and measured
curves collapse on each other when $T/\Delta_{a}$ is in the immediate
neighborhood of 1: thus the apparent collapse is a reminder that our
experimental conditions are good only for probing that part of the phase
diagram wherein the field has a weak influence on the strongly
exchanged-coupled AFM-like\ state. It is worth mentioning that such a
collapsing feature is reflected also, by virtue of Eq.
\ref{Eq-C-M-relations(b)}, in the low-temperature $M$($T,H$)
curves.\cite{Canfield-GdNi2B2C,03-GdNi2B2C-singlecrystal-1}%

\begin{figure}
[th]
\begin{center}
\includegraphics[
natheight=4.369900in,
natwidth=3.257800in,
height=4.3699in,
width=3.2578in
]%
{D:/Spin-Wave/graphics/RNi2B2C-Field-Spheat-Gd-Fit__3.pdf}%
\caption{(Color online) Log-Log plot of $C_{mag}$ versus the normalized
temperature ($T/\Delta$) under different applied fields. $\Delta_{a}$\textrm{
}represent the zero-field energy gap of Eq. \ref{Eq-gap-AFM}. Symbols denote
measurements while solid lines represent the calculated $C_{mag}(T,H)$ based
on Eq. \ref{Eq-Cm-AFM-lowH}. The zero-field $C_{mag}(T,H)$ curve (measured
down to 0.5 K) was fitted with Eq. \ref{Eq-Cm-AFM-lowH}. The obtained
parameters were fed into Eq. \ref{Eq-Cm-AFM-lowH} and there from the
theoretical $C_{mag}(T,H)$ curves for field up to 20 kOe were calculated with
no adjustable parameters (see text).}%
\label{Fig-Gd-Cfit}%
\end{center}
\end{figure}

Figure \ref{Fig-Gd-Ctot} shows also that the $H-$evolution of the
low-temperature $C_{mag}$($T,H_{\parallel c}$) curves is similar to that
of\ $C_{mag}$($T,H_{\parallel a}$) ones; the only difference is that all
values of $H_{\mathrm{\parallel}a}$ are lower than $H_{R}^{\mathrm{\parallel
}a}(T<T_{R})$ boundary while, in contrast, some of the applied $H_{\parallel
c}$ values are higher than $H_{R}^{\mathrm{\parallel}c}(T<T_{R})$ (see above);
then it is no surprise that the presence of the $T_{R}(H_{\parallel c})$-event
in $C_{mag}(T,H_{\parallel c})$ is more pronounced than in $C_{mag}%
(T,H_{\parallel a})$; in fact there is no manifestation of the reorientation
event in the isofield $C_{mag}(T,H_{\parallel c}>$32kOe$)$\ curves while, in
contrast, for $H\mathrm{\parallel}a$, the\ presence of the reorientation event
is evident in all applied field up to the maximum 80 kOe (see insets of Fig.
\ref{Fig-Gd-Ctot}). It is noted that the reentrant feature of the
$H_{R}^{\mathrm{\parallel}a}(T)$ curve within the neighborhood of 11 K is well
evident in the $C_{mag}(T,H_{\parallel a})$ curves: as $H\mathrm{\parallel}a$
is increased, the peak associated with this event moves first to higher
temperature but reverts to a decreasing tendency when $H_{\mathrm{\parallel}%
a}$ reaches values higher than 30 kOe.

Within $T<$ 5 K and $H_{\mathrm{\parallel}a}<H_{R}(T)$\ range, $(\partial
S_{mag}/\partial H)_{T}$, $(\partial C_{mag}/\partial H)_{T}$, and $(\partial
M/\partial T)_{H}$ [Ref. \cite{Canfield-GdNi2B2C,03-GdNi2B2C-singlecrystal-1}]
are weak but positive. Similar features are evident for the
$H\mathrm{\parallel}c$ case. When $H_{\mathrm{\parallel}c}\rightarrow
H_{R}(T)$, $C_{mag}(T,H_{\mathrm{\parallel}c})$ is observed to increase,
reaching a maximum at the phase boundary. A further increase in
$H_{\mathrm{\parallel}c}>H_{R}(T<T_{R})$ leads to $(\partial C_{mag}/\partial
H)_{T}$ $<0$ indicating a decrease in the entropy and as such an increase in
the ordered component along the $c$ axis. For $T>T_{R}$, both $(\partial
M/\partial T)_{H}$ [Refs.\cite{Canfield-GdNi2B2C,03-GdNi2B2C-singlecrystal-1}]
and $(\partial C_{mag}/\partial H)_{T}$ are weak and positive for $H<$30 kOe
but negative for 30 kOe$<H<H_{sat}$.

\paragraph{Model calculation of $C_{mag}(T,H)$}%

\begin{figure}
[th]
\begin{center}
\includegraphics[
natheight=4.719300in,
natwidth=3.456600in,
height=4.7193in,
width=3.4566in
]%
{D:/Spin-Wave/graphics/RNi2B2C-Field-Spheat-Gd-Model-Cal__4.pdf}%
\caption{Comparison of the calculated (solid lines) and measured (symbol)
field-dependent magnetic specific heat of GdNi$_{2}$B$_{2}$C for (a) $H\Vert
a$ $axis$ and (b) $H\Vert c$ $axis$. For ease of visualization, the successive
curves are displaced upwards with the same amount of vertical shift (10
J/moleK). The calculations (with no adjustable parameters) are based on the
model proposed by Jensen and
Rotter\cite{Jensen08-GdNi2B2C-magnetoelstic-paradox} \ (see text).}%
\label{Fig-Gd-CmodelCal}%
\end{center}
\end{figure}

Figure \ref{Fig-Gd-CmodelCal} compares the\ measured magnetic specific heat of
\textrm{GdNi}$_{\text{2}}$\textrm{B}$_{\text{2}}$\textrm{C} with the model
calculation of Jensen and Rotter (see
\S \ II.B).\cite{Jensen08-GdNi2B2C-magnetoelstic-paradox} It is assuring to
notice that, even though there are no adjustable parameters in these
calculations, the model is able to reproduce the main features of the measured
$C_{mag}(2$ K$<T<T_{N}$) for both $H_{\mathrm{\parallel}a}\leq80$ kOe and
$H_{\mathrm{\parallel}c}\leq80$ kOe [see, respectively, Figs.
\ref{Fig-Gd-CmodelCal}(a) and (b)]. The following three achievements of the
model calculations should be highlighted: (i) the calculated magnitude of the
steps at both $T_{R}$ and $T_{N}$ compare favorably with the measured values;
(ii) the surge of an anisotropy for the \textit{spherical} $^{7/2}S$
Gd-moments even at temperatures as high as 20 K is well accounted for; and
(iii) the reorientation process at $T_{R}$ (along both field orientations) is
shown to be a consequence of the joint action of exchange interactions and
dipolar forces even though the energy of the former is at least five times
larger than that of the latter: isotropic bilinear interaction, alone by
themselves, do not lead to any reorientation
processes.\cite{Blanco-Gd-Amplitude-modulated,Bouvier-Gd-amplitude-modulated,rotter01-AM-Gd-compounds}
Evidently in spite of the above-mentioned successes, this mean-field model
calculation are not expected to account for the magnetic features of the
specific heat within the very low-temperatures or in the neighborhood of
$T_{N}$ (the critical region).

\subsubsection{\textrm{HoNi}$_{\text{2}}$\textrm{B}$_{\text{2}}$\textrm{C}}

The zero-field magnetic ground structure of \textrm{HoNi}$_{\text{2}}%
$\textrm{B}$_{\text{2}}$\textrm{C} is an AFM state\ wherein the ferromagnetic
basal planes are piled along the $c$ axis forming an AFM configuration
$\nearrow\swarrow$ with $q$ = (0,0,1).\cite{Lynn97-RNi2B2C-mag-xrd} On
applying a field along the easy (1,1,0) axis, the phase diagram of
\textrm{HoNi}$_{\text{2}}$\textrm{B}$_{\text{2}}$\textrm{C} shows a succession
of field-induced metamagnetic
phases,\cite{Canfield97-Er-Ho-Dy-Tb-HT-diagram,Detlefs00-Ho-Field-neutronDiff,Campbell-Ho-HT-Diagram}
transforming from $\nearrow\swarrow\nearrow$ into: $\nearrow\nearrow\swarrow$
at $H_{1}$, $\nearrow\nearrow\searrow$ at $H_{2}$, and a saturated
$\nearrow\nearrow\nearrow$\ state at $H_{sat}$. On the other hand, if the
field is applied within the $ab$ plane but at an angle $\theta$ away from the
easy axis, then the strong CEF-induced anisotropic character of \textrm{HoNi}%
$_{\text{2}}$\textrm{B}$_{\text{2}}$\textrm{C} becomes evident and as a result
restricts the moments to be along only the
$<$%
110%
$>$
directions: as shown in the lower inset of Fig.\ \ref{Fig-Ho-Ctot}, an
increase in $\theta$ induces an increase in both $H_{1}$ [expressed as
4.1/cos($\theta$) kOe] and $H_{sat}$ [related by 6.6/sin(45-$\theta$) kOe] but
a decrease in $H_{2}$ [described by 8.4/cos(45-$\theta$)
kOe].\cite{Canfield97-Ho-angularMetamagnet}%

\begin{figure}
[th]
\begin{center}
\includegraphics[
natheight=4.528200in,
natwidth=3.107300in,
height=4.5282in,
width=3.1073in
]%
{D:/Spin-Wave/graphics/RNi2b2C-Field-Spheat-Ho-tot__5.pdf}%
\caption{(Color online) Thermal evolution of the magnetic specific heats of
\textrm{HoNi}$_{\text{2}}$\textrm{B}$_{\text{2}}$\textrm{C} under various
applied fields.\cite{04-Magnetocaloric} For clarity, the $C_{mag}(T,H\leq10$
kOe$)$ curves are plotted in the lower panel while the $C_{mag}(T,H\geq10$
kOe$)$ curves are plotted in the upper panel. The lower\ inset shows the
angular dependence of the critical fields\ (adapted from Ref.
\cite{Canfield97-Ho-angularMetamagnet}); the horizontal dashed lines represent
the applied fields. The upper inset shows the field-dependence of the magnetic
specific heat at 2 K: $C_{mag}(T,H)$ within the same metamagnetic state is
almost the same; this feature is reflected also in the $M(2K,H)$ isotherms
reported in Ref. \cite{Canfield97-Ho-angularMetamagnet} (see text).}%
\label{Fig-Ho-Ctot}%
\end{center}
\end{figure}

Figure \ref{Fig-Ho-Ctot} shows the thermal evolution of the isofield
$C_{mag}(T,H)$ of \textrm{HoNi}$_{\text{2}}$\textrm{B}$_{\text{2}}$\textrm{C}
for various fields, all applied along the (100) axis ($\theta$=45$%
{{}^\circ}%
$); its overall field-dependence can be comprehended in the light of Eq.
\ref{Eq-C-M-relations(b)}, the $M(T,H)$ curves, and the features of the
magnetic phase diagrams. One of the most remarkable features of these
$C_{mag}(T,H)$ curves is the observation that their field-dependence (the
upper inset of Fig. \ref{Fig-Ho-Ctot}) is reminiscent of the field-dependence
of the isothermal $M(T,H)$: as an example, for $H$ $\leq$ $H_{1}%
$,\cite{Canfield-HoNi2B2C-M-Cp} $\left(  \partial^{2}M/\partial^{2}T\right)
_{H}>0$ and concomitantly $C_{mag}(T,H)$ (see Fig. \ref{Fig-Ho-Ctot})
increases with $H$. Furthermore, $C_{mag}(T,H_{1}<H<H_{2})$ is relatively much
higher but $C_{mag}(T,H>H_{2})$ decreases monotonically with the field as it
it approaches the forced saturated $\nearrow\nearrow\nearrow$ state.

Based on the $H-\theta$ phase diagram (lower inset of Fig. \ref{Fig-Ho-Ctot}),
it becomes clear that among the various $C_{mag}(T,H)$ curves, the only
candidates to be contrasted with the spin wave analysis\ of \S \ II.A.1 are
those measured within the $H$
$<$%
$H_{1}$ range, namely $H_{\parallel\mathrm{a}}=$ 0 and 2.5 kOe wherein the
low-$T$ collinear AFM $\nearrow\swarrow$ state is
established.\cite{Canfield97-Ho-angularMetamagnet} Fig. \ref{Fig-Ho-Cfit}
shows the satisfactorily fit of $C_{mag}(T,H=0)$ and the best fit parameters
are: $\Delta=7.7\pm0.3$ K and $D=$4.6$\pm0.2$ K. These values are in agreement
with our earlier results.\cite{04-Magnetocaloric} Based on Eq.
\ref{Eq-Cm-AFM-lowH}, the thermal evolution of the $C_{mag}(T,H_{\parallel
\mathrm{a}}=$ 2.5 kOe) should be obtainable form these $\Delta$ and $D$ and
the value of the field component along the (110) direction; this is indeed the
case as can be convincingly observed in Fig.\ \ref{Fig-Ho-Cfit}.
\begin{figure}
[th]
\begin{center}
\includegraphics[
natheight=4.687300in,
natwidth=3.418600in,
height=4.6873in,
width=3.4186in
]%
{D:/Spin-Wave/graphics/RNi2B2C-Field-Spheat-Ho-Fit__6.pdf}%
\caption{(Color online) A log-log plot of the isothermal $C_{mag}(T/\Delta
_{a},H$ ) curves of \textrm{HoNi}$_{2}$\textrm{B}$_{2}$\textrm{C. }$\Delta
_{a}$\textrm{ }represent the zero-field energy gap of Eq. \ref{Eq-gap-AFM}.
For clarity, the $C_{mag}(T/\Delta_{a},H\leq10$ kOe$)$ curves are plotted in
the lower panel while those of $C_{mag}(T/\Delta_{a},H\geq10$ kOe$)$ curves
are plotted in the upper panel. The solid line represents the non-linear least
square fit of the zero-field measured curve to Eq. \ref{Eq-Cm-AFM-lowH} (see
text).}%
\label{Fig-Ho-Cfit}%
\end{center}
\end{figure}

\subsubsection{\textrm{ErNi}$_{\text{2}}$\textrm{B}$_{\text{2}}$\textrm{C}}

The magnetic phase diagram of \textrm{ErNi}$_{\text{2}}$\textrm{B}$_{\text{2}%
}$\textrm{C} is shown in the inset of Fig. \ref{Fig-Er-Ctot}%
:\cite{Budko00-ErNi2B2C-angularHTdiagram} in zero field, this compound
superconducts at 10.5 K and orders magnetically at $T_{N}$ =6.4 K into an
incommensurate modulated AFM\ structure with $q$=(0.55,0,0) and moments
pointing along the $b$ axis.\cite{Lynn97-RNi2B2C-mag-xrd} Below $T_{WFM}$ =2.2
K, a weak ferromagnetic (WFM) state\cite{Canfield96-Er-WF-HT-diagram} emerges
together with an equal-amplitude, squared-up
state.\cite{Kawano02-ErNi2B2C-neutron,Choi01-ErNi2B2C-neutron} A series of
field-induced metamagnetic transformations appears when a field ($H$%
\textrm{$<$}$H_{sat}$) is applied along the easy axis
(0,1,0):\cite{Campbell00-Er-HT-diagram,Budko00-ErNi2B2C-angularHTdiagram,Canfield97-Er-Ho-Dy-Tb-HT-diagram,02-ErNi2B2C-magnetostriction}
at 2 K, three metamagnetic transformations occur at 4, 11, and 20 kOe. For
$H>H_{sat}$, the paramagnetic saturated state is stabilized.%
\begin{figure}
[th]
\begin{center}
\includegraphics[
natheight=4.687300in,
natwidth=3.191200in,
height=4.6873in,
width=3.1912in
]%
{D:/Spin-Wave/graphics/RNi2B2C-Field-Spheat-Er-tot__7.pdf}%
\caption{(Color online) Thermal evolution of $C_{mag}(T,H$) of \textrm{ErNi}%
$_{2}$\textrm{B}$_{2}$\textrm{C} at various magnetic fields. The inset shows
the $H-T$ phase diagram (adapted from Ref.
\cite{Budko00-ErNi2B2C-angularHTdiagram}): the thick dashed line represent the
$H_{c2}(T)$ $-$superconductivity$-$ curve while the solid lines represent the
various magnetic transformations. The horizontal dotted lines represent the
magnetic fields that were used during this study.}%
\label{Fig-Er-Ctot}%
\end{center}
\end{figure}

To investigate the $H$- and $T$-dependence of $C_{mag}(T,H)$ across such a
phase diagram, we carried out a series of isofield measurements with
$H\Vert(010)$. The resulting $C_{mag}(T,H_{\Vert a})$ curves are shown in
Figs. \ref{Fig-Er-Ctot}. The thermal evolution of the zero-field entropy (not
shown) suggests that the lowest four level are fully populated above 10 K;
this result is in agreement with the findings of Gasser \textit{et
al.}\cite{Gasser96-CEF-HoErTmNi2B2C} that the electronic ground state is a
doublet which is separated from the immediate excited state (also a doublet)
by 0.6 $\sim$ 0.7 meV (7$\sim$8 K). For $H<H_{1}$ ( $T<T_{WFM}$), an increase
in $H$ induces an increase in $S_{mag}(H)$ and, based on Eq.
\ref{Eq-C-M-relations(a)} and the experimental
results,\cite{Cho95-ErNi2B2C-ansitropy} $(\partial M/\partial T)_{H}>0$
(considerations should be made for the presence of the superconductivity below
$H_{c2}$ and the surge of weak ferromagnetic state below $T_{WFM}$).
\begin{figure}
[th]
\begin{center}
\includegraphics[
natheight=4.686400in,
natwidth=3.266400in,
height=4.6864in,
width=3.2664in
]%
{D:/Spin-Wave/graphics/RNi2B2C-Field-Spheat-Er-fit__8.pdf}%
\caption{(Color online) A$\ $log-log plot of the $C_{M}(T,H)$ curves of
\textrm{ErNi}$_{2}$\textrm{B}$_{2}$\textrm{C} at various applied magnetic
fields. The magnetic specific heat (symbol) are compared with the theoretical
calculation (lines) based on Eq. \ref{Eq-Cm-AFM-lowH} (see text). The
calculated curves of $H$=0, 5, 10 kOe are different from each other only at
$T<\Delta$.}%
\label{Fig-Er-Cfit}%
\end{center}
\end{figure}

For $T<$ $T_{WFM}(H<H_{1})$, the zero-field magnetic sate is approximated as
an AFM structure (see \S \ II.A ); accordingly, the measured $C_{mag}%
(T<T_{WFM},H<H_{1})$ curves are confronted with Eq. \ref{Eq-Cm-AFM-lowH}. Fig.
\ref{Fig-Er-Cfit} shows the excellent fit of $C_{mag}(T,H=0$ kOe$)$\ which
gives $\Delta=7.0\pm0.1$ K and $D=$3.0$\pm0.1$ K. Using these parameters, we
calculated the\ field-dependent $C_{mag}(T,H)$\ curves (see Fig.
\ref{Fig-Er-Cfit}). Once more, one observes the collapse of the $C_{mag}%
(T<T_{WFM},H<H_{1})$ curves within the immediate neighborhood of $T_{WFM}^{-}%
$. Due to experimental limitations, we were not able to probe the field
evolution for $T<T_{WFM}$ nor the thermal evolution for $H>$ $H_{sat}$ of
$C_{mag}(T,H)$ of \textrm{ErNi}$_{\text{2}}$\textrm{B}$_{\text{2}}$\textrm{C}.

\section{Discussion and Summary}

The general evolution of $C_{mag}(T,H)$ within the magnetically ordered state
of representative $R$\textrm{Ni}$_{\text{2}}$\textrm{B}$_{\text{2}}$\textrm{C}
($R$=\textrm{Gd, Ho, Er}) is found to reflect faithfully the characteristic
features of their $H-T$ phase diagrams. Three approaches are employed for the
analysis of the evolution of these $C_{mag}(T,H)$ curves: (i) basic
thermodynamical analysis which allows us to relate the evolution of
$C_{mag}(T,H)$ to that of the magnetization measurements: thus permitting a
generalization to field and temperature ranges beyond the limitation of our
experimental set ups. (ii) the linearized spin-wave analysis which allows us
to investigate the low-temperature, low-field range; and (ii) the model
calculation based on which we are able to probe the higher temperature region,
a region which is not accessible for spin-wave analysis.

One of the characteristic features of the magnetic phase diagrams is the
stair-like behavior observed in the magnetization isotherms: it is assuring
that this feature is manifested also in the $C_{mag}(T,H)$ case: all the
$C_{mag}(T,H)$ curves within the same metamagnetic mode do collapse on each
other (see e.g. Fig. \ref{Fig-Ho-Ctot}). This feature is related to the
influence of the elementary magnetic excitations on the thermodynamical
properties and here, in this work, we discuss this influence in terms of the
linearized spin-wave model (see the theory in \S \ II.A and its confrontation
with the measured $C_{mag}(T,H)$ curves in \S \ IV.C)

Based on the analysis of \S \ IV one is able to delineate the low-field,
low-temperature range of the $H-T$ phase diagrams wherein the linearized
magnon contribution is found to describe satisfactory the experimental
results: for \textrm{GdNi}$_{\text{2}}$\textrm{B}$_{\text{2}}$\textrm{C}, it
is the $H<H_{R}(T)$ range wherein the sine-modulated state is squared-up; for
\textrm{ErNi}$_{\text{2}}$\textrm{B}$_{\text{2}}$\textrm{C} it encompasses the
$H<H_{1}$($T<$ $T_{WFM})$ range wherein the squared-up AFM-like state is
established; and finally for \textrm{HoNi}$_{\text{2}}$\textrm{B}$_{\text{2}}%
$\textrm{C}, it is the $H<H_{1}$ ($T<$ $T_{N})$ range wherein the collinear
AFM structure is established. In all these magnetic states, the dispersion
relation is taken to be given by Eq. \ref{Eq-disp} and the magnetic specific
heat is expressed by Eq. \ref{Eq-Cm-AFM-lowH}: the satisfactorily agreements
between the calculated and measured $C_{mag}(T,H)$ justifies the assumptions
considered in this model.

The values of the fit parameters evolves reasonably well across the studied
compounds: as an example, the gap parameter of \textrm{GdNi}$_{\text{2}}%
$\textrm{B}$_{\text{2}}$\textrm{C} ($\Delta$ $=2.9$ K) is much smaller than
the corresponding values of \textrm{HoNi}$_{\text{2}}$\textrm{B}$_{\text{2}}%
$\textrm{C} ($\Delta$ $=7.7$ K) and \textrm{ErNi}$_{\text{2}}$\textrm{B}%
$_{\text{2}}$\textrm{C} ($\Delta$ $=7.0$ K); based on Eq. \ref{Eq-gap-AFM},
such a result does agree with the well-established fact that the anisotropic
field for \textrm{GdNi}$_{\text{2}}$\textrm{B}$_{\text{2}}$\textrm{C} is
extremely small. On the other hand, the stiffness constant of \textrm{GdNi}%
$_{\text{2}}$\textrm{B}$_{\text{2}}$\textrm{C} ($D=5.5$ K) is greater than
that of \textrm{HoNi}$_{\text{2}}$\textrm{B}$_{\text{2}}$\textrm{C} ($D$
$=4.6$ K) and \textrm{ErNi}$_{\text{2}}$\textrm{B}$_{\text{2}}$\textrm{C} ($D$
$=3.0$ K); based on Eq. \ref{Eq-Stiff-AFM}, this is related to the fact that
the effective exchange couplings of \textrm{GdNi}$_{\text{2}}$\textrm{B}%
$_{\text{2}}$\textrm{C} (proportional to the de Gennes factor) are the
strongest. It is recalled that no direct scaling with the de Gennes factor
should be expected since the $D$ term of Eq. \ref{Eq-Stiff-AFM} contains also
$H_{a}$.

As the investigated compounds are good representatives of the other magnetic
borocarbides, then it is expected that the above mentioned characteristic
step-like behavior should be manifested also in other magnetic $R$%
\textrm{Ni}$_{\text{2}}$\textrm{B}$_{\text{2}}$\textrm{C} compounds: indeed,
the $M(T,H)$\ isotherms of \textrm{TbNi}$_{\text{2}}$\textrm{B}$_{\text{2}}%
$\textrm{C} (having similar\ anisotropic features as those of \textrm{ErNi}%
$_{\text{2}}$\textrm{B}$_{\text{2}}$\textrm{C}) manifest such a stair-like
feature.\cite{Canfield97-Er-Ho-Dy-Tb-HT-diagram,Cho96-TbNi2B2C-anistropy-WF,07-TbNi2B2C}
Then, based on Eq. \ref{Eq-C-M-relations(b)}, its $C_{mag}(T,H)$ features
should be similar to the those of the studied compounds. It is emphasized that
the manifestation of such stair-like features is a more general property since
it is manifested, not only in the $M(T,H)$ isotherms, but also in several
other thermodynamical quantities (see \S \ II.B). This generality suggests
that, although other models have been applied to analyze these stair-like
features in, e.g., \textrm{HoNi}$_{\text{2}}$\textrm{B}$_{\text{2}}$\textrm{C}
[Ref. \cite{Kalatsky98-clock-model,Amici-Thalmeier98-HoNi2B2C}] and
\textrm{ErNi}$_{\text{2}}$\textrm{B}$_{\text{2}}$\textrm{C}%
,\cite{Jensen02-mode-FM-ErNi2B2C} an analysis in terms of the spin-wave model
is shown to be extremely useful for the description of the low-temperature
thermodynamical properties of these magnetic compounds.

\begin{acknowledgments}
We acknowledge the partial financial support from the Brazilian agencies CNPq
(485058/2006-5) and Faperj (E-26/171.343/2005) and the Austrian Science
Foundation (FWF) P16250.
\end{acknowledgments}

\bibliographystyle{apsrev}
\bibliography{Borocarbides,Crystalograph,Intermetallic,Mag-classic,Massalami,ND-RepAnalysis,Nuclear-hyperfine-MES,To-Be-Published}

\appendix

\section{The high-field limit ($H>H_{sat}$)}

For the induced FM state, we assume that the action of both the anisotropic
field and exchange couplings is the similar to the case of the low-field limit
(\S \ II.A) except that here the field is strong enough to overturn the
antiparallel spins: all the spins are oriented along the easy-axis directions;
e.g. for the case of \textrm{HoNi}$_{\text{2}}$\textrm{B}$_{\text{2}}%
$\textrm{C}, it is one of the four
$<$%
110%
$>$
directions that is nearest to the field. In the long wave limit, the
dispersion relation simplifies to:
\begin{equation}
\hbar\omega_{k}=\Delta_{f}+c_{1}(k_{x}^{2}+k_{y}^{2})-c_{2}k_{z}^{2},
\label{Eq-disp-FM}%
\end{equation}
where, assuming a weaker orthorhombic distortion (a$\approx$b):%
\begin{align*}
c_{1}  &  =2Sa^{2}(\mathcal{J}_{0}-\mathcal{J}_{1})\\
c_{2}  &  =2Sc^{2}\mathcal{J}_{1},
\end{align*}
and the energy gap ($k=0$):%
\begin{equation}
\Delta_{f}=2g\mu_{B}Hcos(\theta)/k_{B}. \label{Eq-gap-FM}%
\end{equation}

Thus, according to the above assumptions $-$ involving the anisotropic field
and exchange couplings $-$ the spin-wave parameters of the saturated FM state
are different from the standard (spontaneous) FM state: in the latter case the
gap includes a 2$H_{a}$ term and the sign of the $c_{2}$\ term in Eq.
\ref{Eq-disp-FM} is positive. Applying standard procedure, the magnon
contribution is given as:%
\begin{equation}
C_{mag}(T)=\frac{R\Delta_{f}^{3/2}}{2^{3/2}\pi^{2}D_{f}^{3/2}T}\sum
_{m=1}^{\infty}\left[  K_{2}(m\Delta_{f}/T)\right]  , \label{Eq-Cm-FM-highH}%
\end{equation}
where
\begin{equation}
D_{f}=2S(\mathcal{J}_{0}-\mathcal{J}_{1})^{\frac{2}{3}}\mathcal{J}_{1}%
^{\frac{1}{3}}. \label{Eq-Stiff-FM}%
\end{equation}

For $T$ $<\Delta_{f}$, Eq. \ref{Eq-Cm-FM-highH} reduces to $R\Delta
exp(-\Delta/T)(\pi^{3/2}D_{f}^{3/2}T^{1/2})$ while for $T$ $>>\Delta_{f}$ it
gives the well-known $T^{3/2}$ relation.\cite{Kittel-SpinWave}

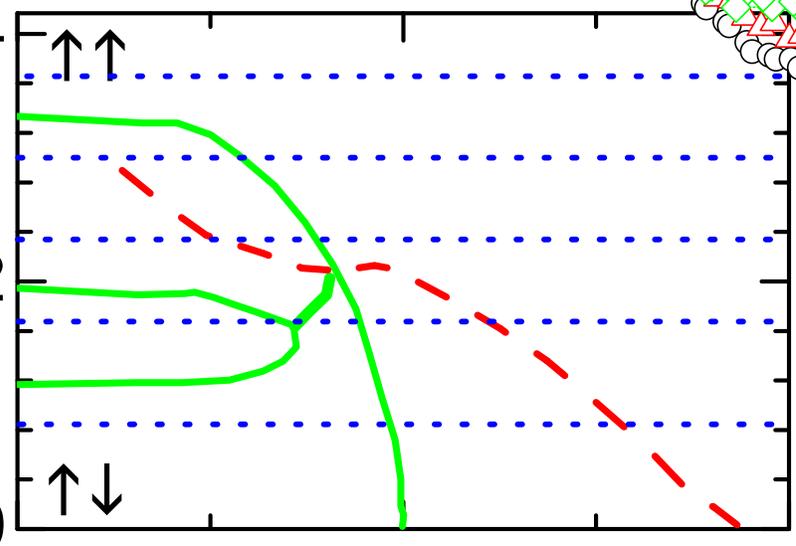
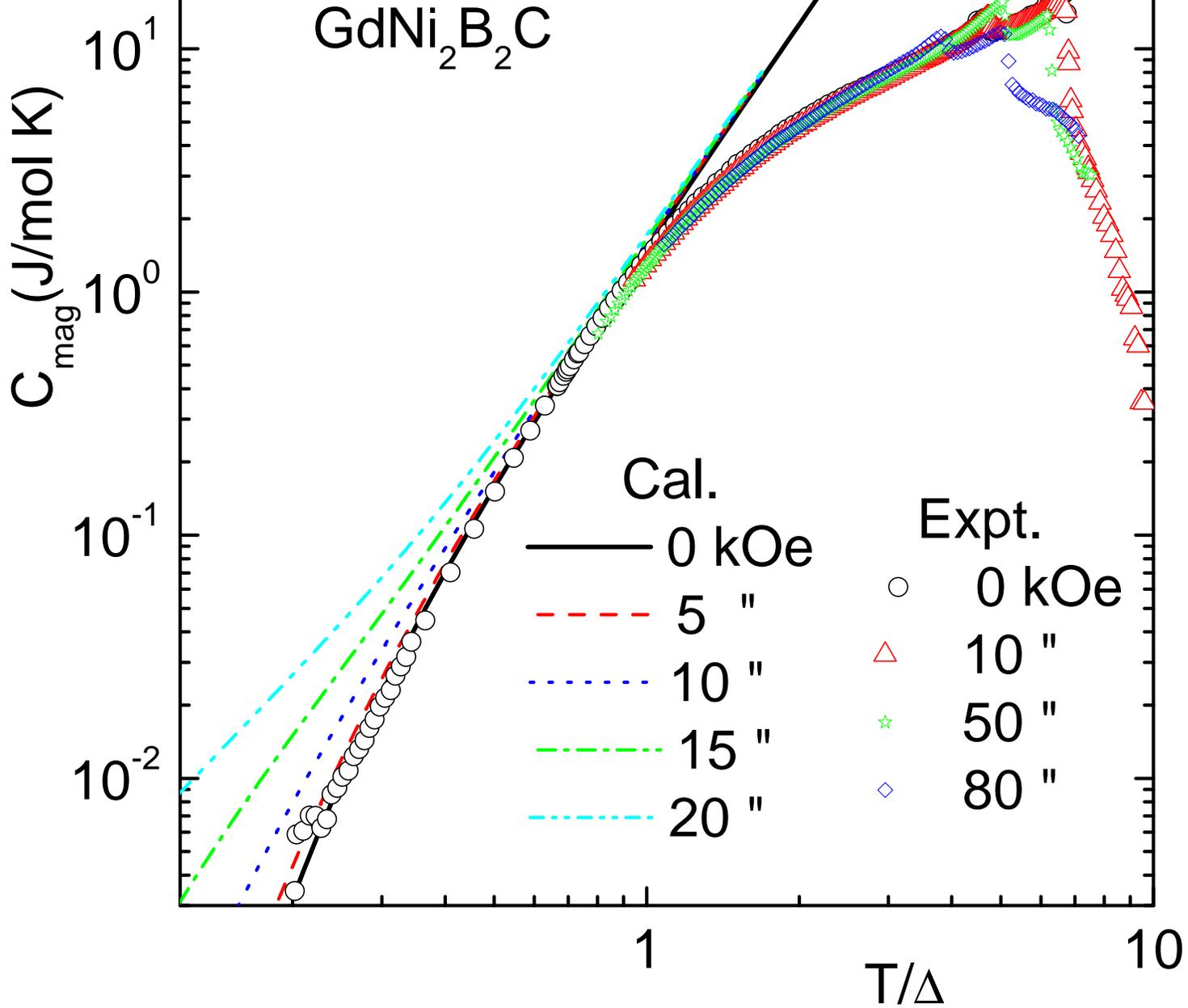
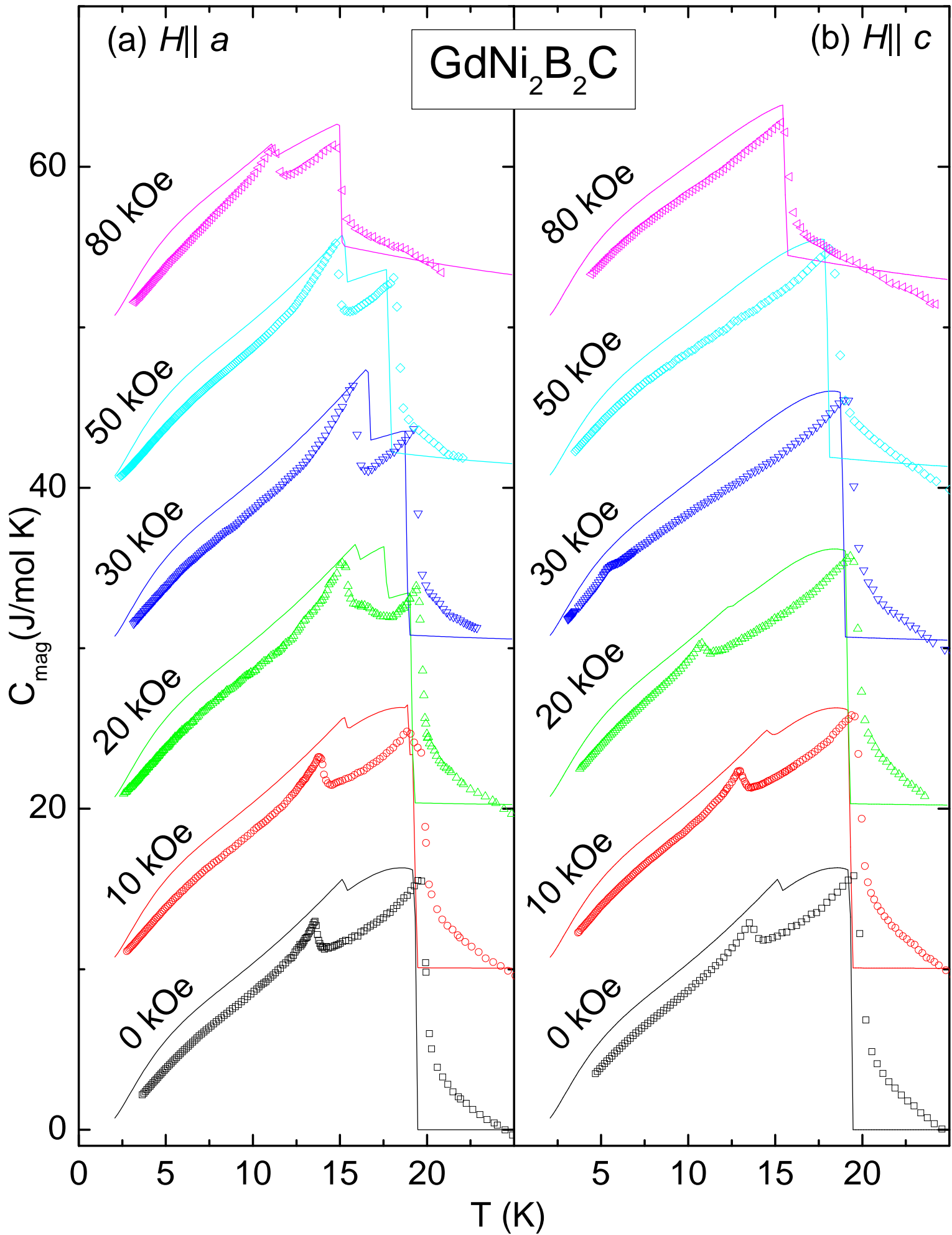
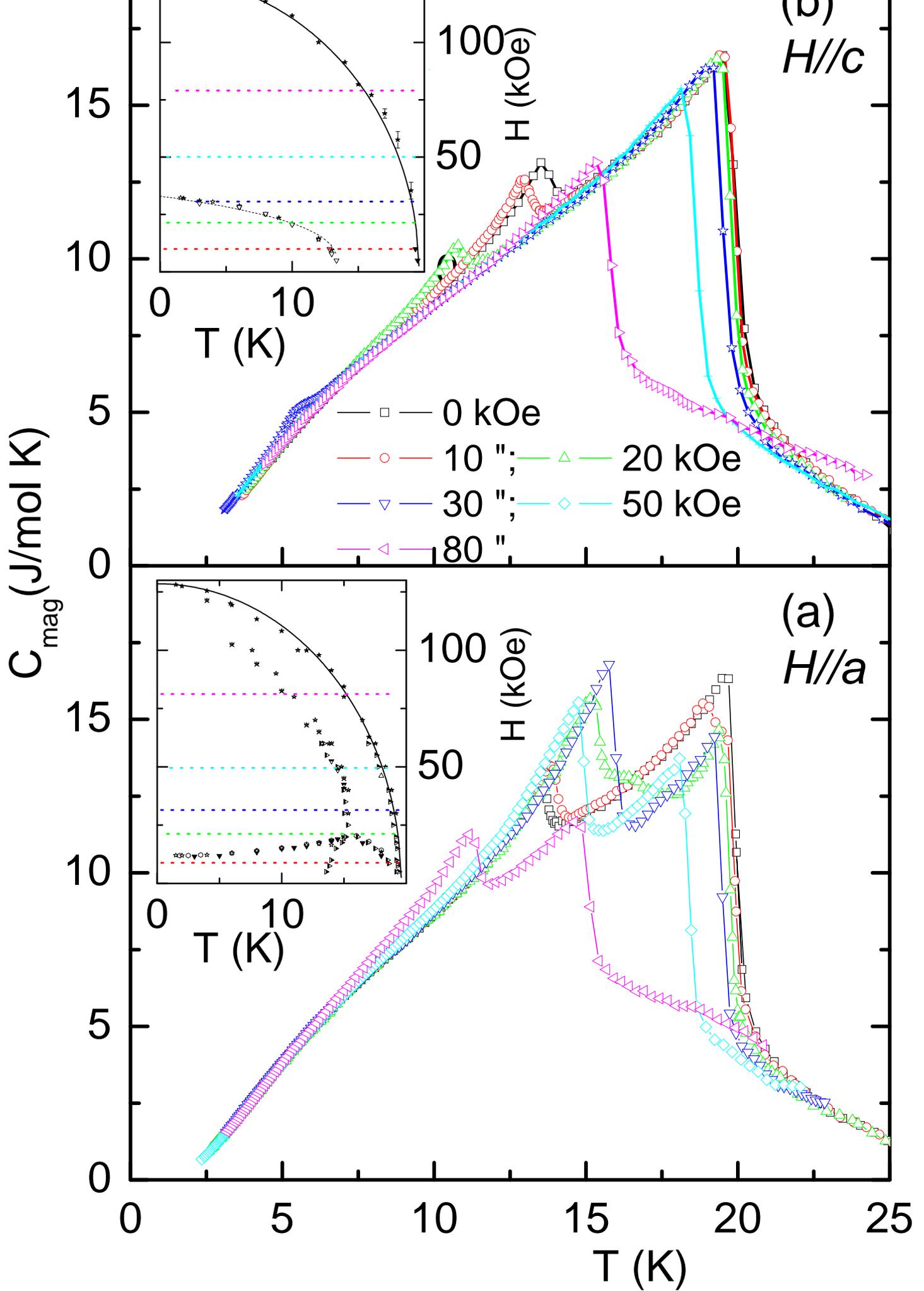
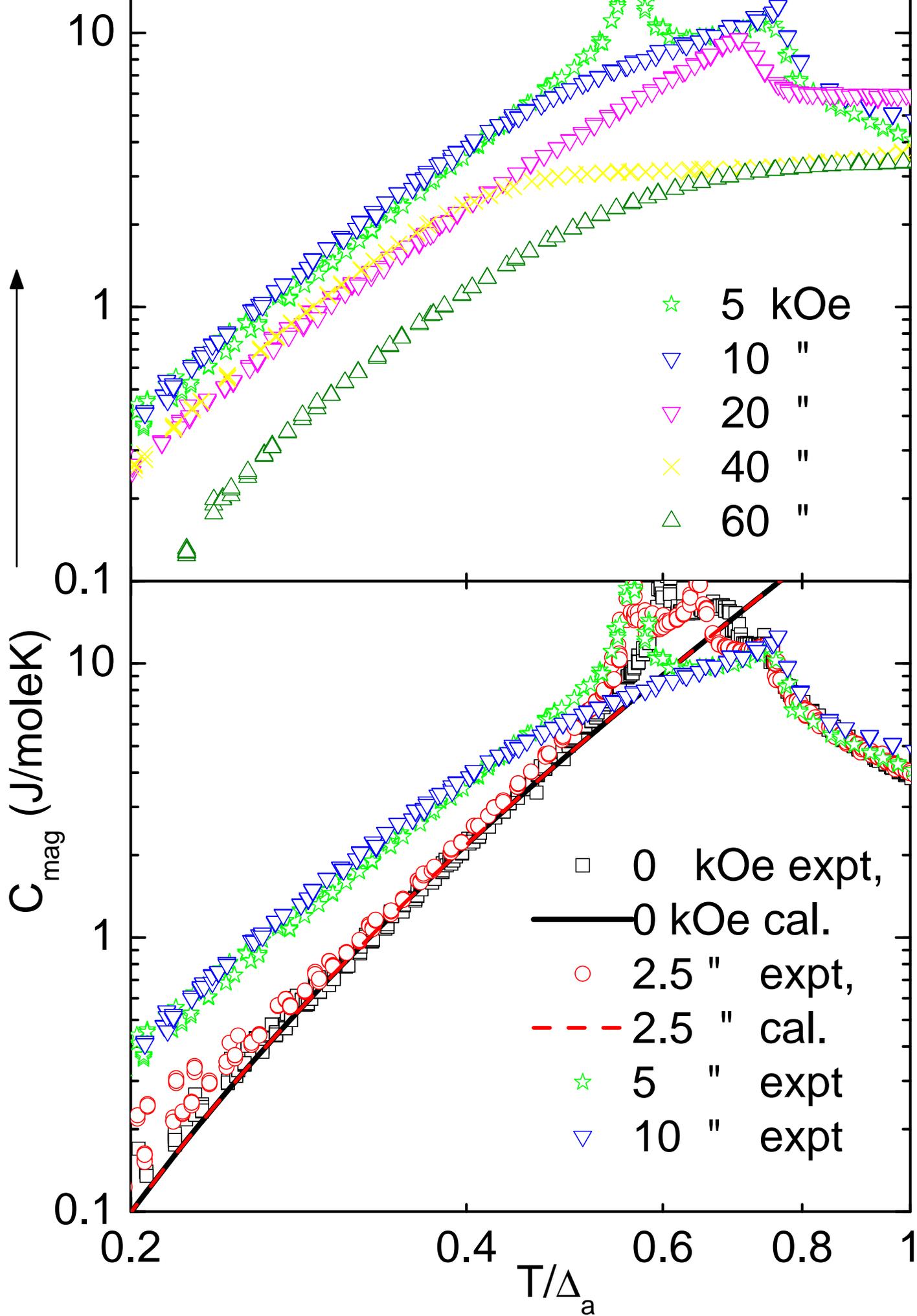
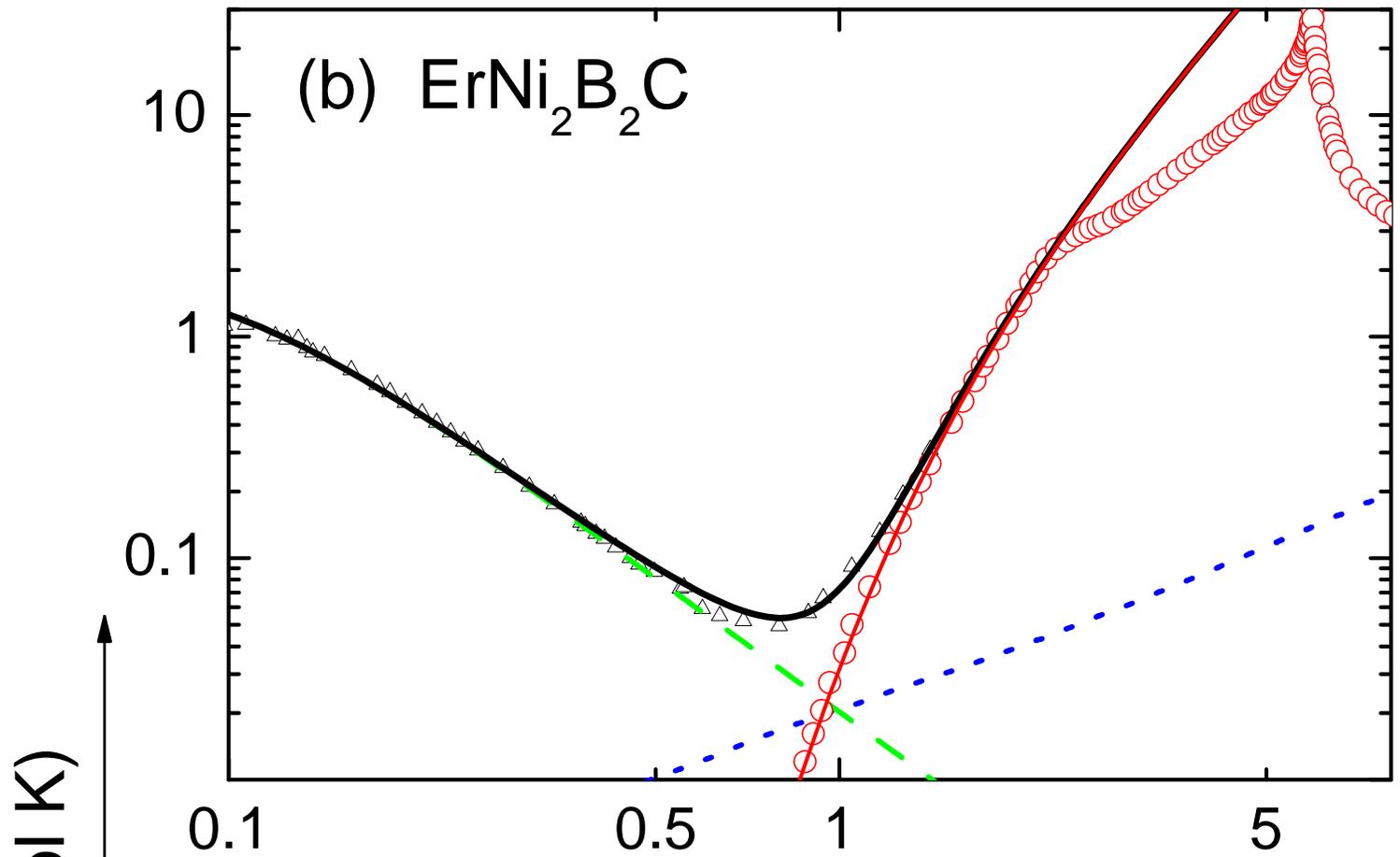
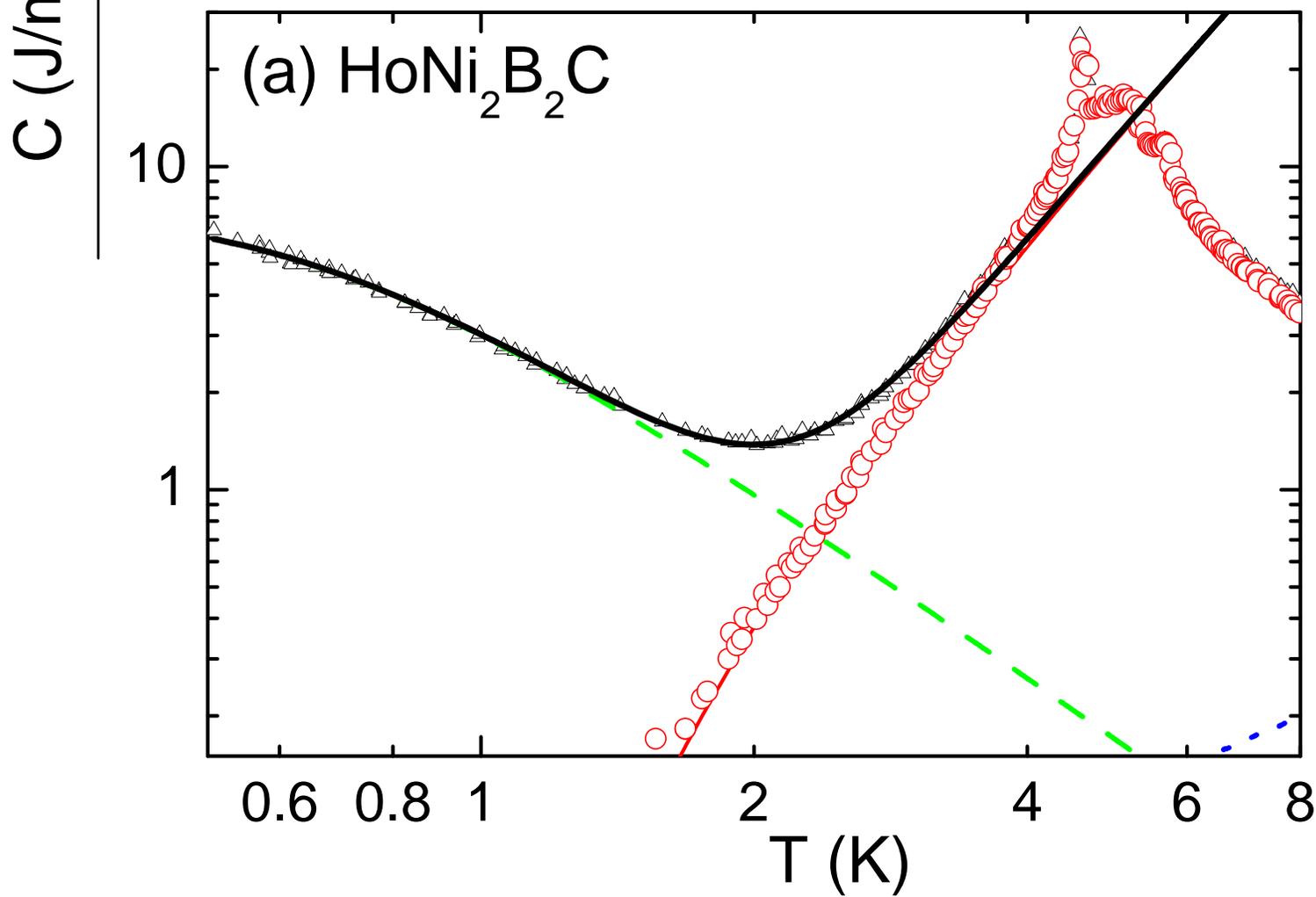
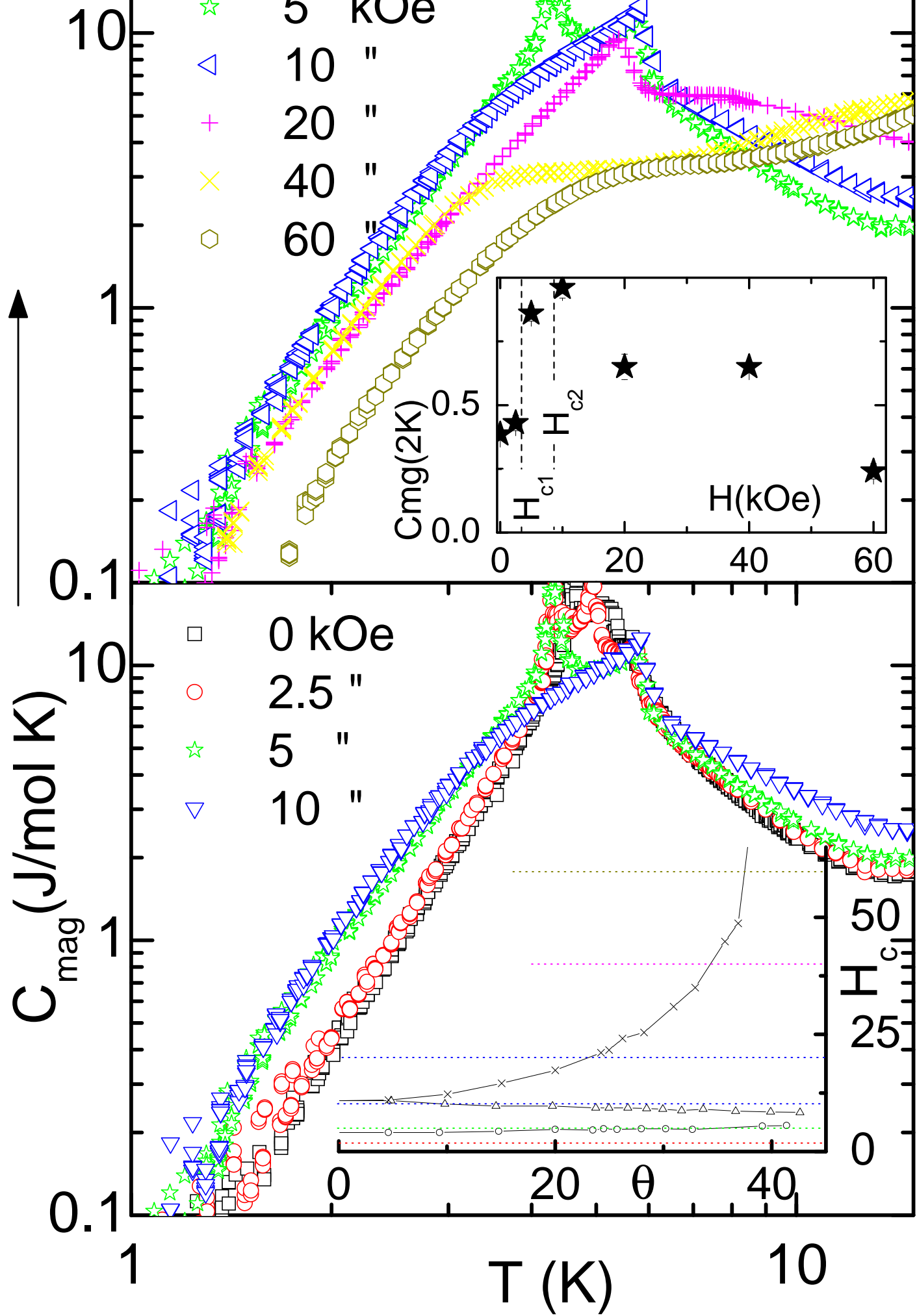
\end{document}